\documentclass{article}
\usepackage{graphicx} 

\usepackage[margin = 3.5cm]{geometry}
\usepackage{times}
\usepackage[figurename=Fig.]{caption}
\usepackage{amsmath}
\usepackage{upgreek}
\usepackage{titlesec}

\titleformat{\section}
  {\normalfont\large\bfseries}{\thesection.}{0.25cm}{}

\title{\textbf{Reliable Micro-Transfer Printing Method for Heterogeneous Integration of Lithium Niobate and Semiconductor Thin Films}}
\author{\large Tom Vandekerckhove$^{1,2,*}$, Tom Vanackere$^{1,2}$, Jasper De Witte$^{1}$, Stijn Cuyvers$^{1}$,\\ 
\large Luis Reis$^{1,2}$, Maximilien Billet$^{1}$, Günther Roelkens$^{1}$, Stéphane Clemmen$^{1,2,3}$,\\
\large and Bart Kuyken$^{1}$\\}
\date{\small{1. Photonics Research Group, INTEC, Ghent University - imec, 9052 Ghent, Belgium}\\
\small{2. OPERA-Photonique CP 194/5, Université Libre de Bruxelles, 1050 Brussels, Belgium}\\
\small{3. Laboratoire d'Information Quantique, Université Libre de Bruxelles, 1050 Brussels, Belgium}\\
\vspace{2mm}
$^{*}$tom.vandekerckhove@ugent.be}

\begin{document}

\maketitle

\begin{abstract}
High-speed Pockels modulation and second-order nonlinearities are key components in optical systems, but CMOS-compatible platforms like silicon and silicon nitride lack these capabilities. Micro-transfer printing of thin-film lithium niobate offers a solution, but suspending large areas of thin films for long interaction lengths and high-Q resonators is challenging, resulting in a low transfer yield. We present a new source preparation method that enables reliable transfer printing of thin-film lithium niobate. We demonstrate its versatility by successfully applying it to gallium phosphide and silicon, and provide an estimate of the transfer yield by subsequently printing 25 lithium niobate films without fail.
\end{abstract}

\section{Introduction}

Photonic integrated circuits enable the construction of complex optical systems with small footprints in a scalable and controlled way, with applications ranging from telecommunications \cite{wang2020silicon, rahim2021taking} to quantum technologies \cite{wang2020integrated, moody20222022}. Silicon (Si) and silicon nitride (SiN) photonics have established themselves as mature photonic platforms due to their CMOS compatibility, but lack certain key components such as optical gain, fast Pockels modulation and $\chi^{(2)}$ nonlinearities. A promising approach to circumvent these shortcomings is to heterogeneously integrate other materials at a back-end level to remain CMOS-compatible \cite{kaur2021hybrid}. Integrating direct band gap III-V materials brings gain to the integrated circuits \cite{liang2021recent}, enabling fabrication of on-chip lasers \cite{xiang2021high,de2020heterogeneous} and detectors \cite{piels2014low,goyvaerts2020transfer}. Introducing thin-film lithium niobate and other ferro-electric materials gives access to high-speed Pockels modulation \cite{boynton2020heterogeneously, alexander2018nanophotonic} and second order optical nonlinearities \cite{boes2023lithium}.

A mature heterogeneous integration technique consists of die-to-wafer or wafer-to-wafer bonding \cite{ghosh2023wafer, snigirev2023ultrafast}. The wafer containing the device material is bonded on a target wafer, followed by a substrate removal step and device patterning. While lithographic alignment accuracy is obtained in the fabrication of optical devices, the large bonding footprint combined with the required back-end processing makes co-integration of multiple materials challenging.
A less mature but promising heterogeneous integration method for integrated photonics is micro-transfer printing ($\upmu$TP)\cite{vanackere2020micro, li2022photonic, roelkens2022micro}. Devices are pre-fabricated on a source wafer, after which they are picked with a PDMS stamp and printed on the target wafer through kinetically controlled adhesion. While the alignment is determined through the printing process, with state-of-the-art tools providing $\pm 0.5$ $\upmu$m $3 \sigma$ alignment accuracy \cite{roelkens2022micro}, dense co-integration of different materials is possible with limited processing required after printing. Moreover, arrays of devices can be transferred in a single step, giving a high throughput.

To obtain a high transfer yield in micro-transfer printing, the pre-fabricated devices are suspended on the source wafer, only connected at several points to the source substrate. In case of semiconductor optical amplifiers and diodes, the devices are typically several microns thick, limiting their mechanical flexibility. In contrast, for thin films like lithium niobate (LN) or gallium phosphide (GaP), the layers are only several hundreds of nanometers in thickness. At the same time, large areas are desirable to enable long interaction lengths and the definition of larger structures such as racetrack resonators. However, the suspension of large areas of thin films is generally challenging.
In this paper, we developed a new source preparation method to reliably micro-transfer print large areas of thin films based on pillar-like support structures. We discuss the optimization of the method and demonstrate reliable transfer-printing of thin-film lithium niobate. Once optimized, the thin film areas can easily be enlarged without added process development. While optimized with pillar-like structures, we show the method works with other support configurations and give an estimate of the yield by printing 25 lithium niobate thin films without fail onto silicon nitride waveguides. Lastly, we demonstrate the versatility of the method by successfully applying it on large area thin film gallium phosphide and silicon.

\section{Micro-transfer printing and challenges}

\begin{figure}[h]
\centering
\captionsetup{margin=1cm}
\includegraphics[width=\textwidth]{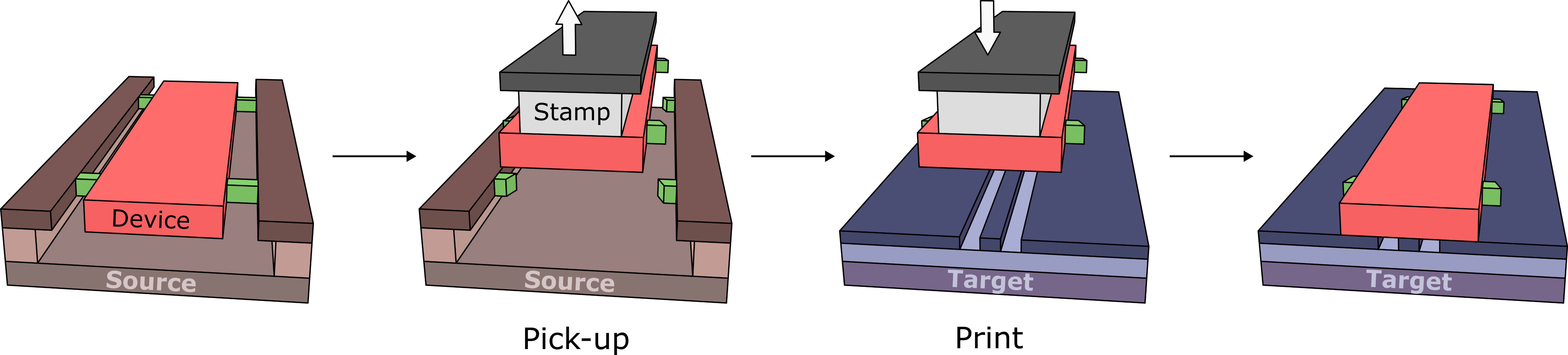}
\caption{Micro-transfer printing concept: a suspended coupon (red) is picked with a PDMS stamp from the source wafer, breaking the tethers (green) in a controlled way, after which it is printed onto the target wafer.}
\label{fig:TPsketch}
\end{figure}

Micro-transfer printing is a pick-and-place method that allows for the heterogeneous integration of many materials, with limited post-processing required. This makes it compatible with CMOS technology as the integration only happens on a back-end level, keeping the front-end process flow unaffected and avoiding any contamination. The micro-transfer printing process is schematically illustrated in Fig.\ref{fig:TPsketch}. 

A source wafer is fabricated where the to-be-integrated devices (red in Fig.\ref{fig:TPsketch}), from now on referred to as coupons, are prepared. These can range from optical amplifiers and detectors, to modulators, but consist of thin-film lithium niobate in the framework of this paper. The layer stack contains a release layer underneath the coupon, which can be selectively etched. Connections, referred to as tethers (green in Fig.\ref{fig:TPsketch}) keep the coupons attached to the wafer. In a last preparation step, the coupons are suspended by etching away the release layer and are ready to be picked. In parallel, the target wafer containing e.g. the photonic circuit is fabricated. An elastomeric PMDS stamp is then used to pick the coupon from the source wafer by breaking the tethers in a controlled manner, and print the device onto the target wafer through direct or adhesive bonding.

\begin{figure}[t]
\centering
\captionsetup{margin=1cm}
\includegraphics[width=\textwidth]{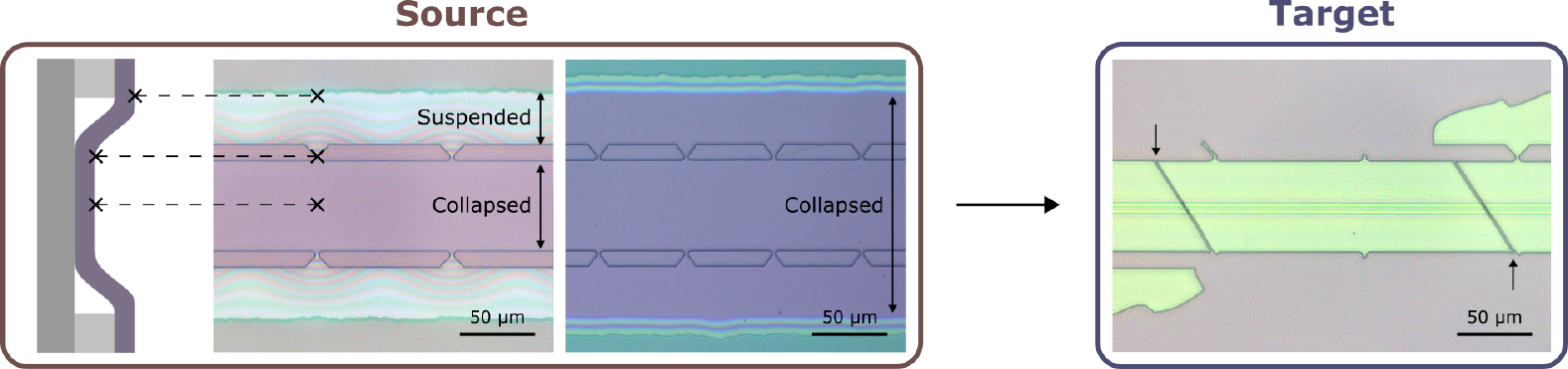}
\caption{Challenges with micro-transfer printing of thin films. Left: Collapsed coupons on the source wafer with a corresponding cross section. Right: Resulting printed LN coupon on a SiN waveguide showing cracks (indicated by arrows) and bad tether-breaking.}
\label{fig:TPchallenges}
\end{figure}

For thin films such as thin-film lithium niobate, the suspension of large areas can be challenging due to their mechanical flexibility. To demonstrate this, we start from a 300 nm x-cut LN/2 $\upmu$m SiO$_2$/LN layer stack (NanoLN) and pattern simple rectangles with hourglass-shaped tethers to the side in the 300 nm LN device layer. The underlying SiO$_2$ layer serves as release layer, which can be selectively etched with hydrofluoric (HF) acid. The release etch is typically optimized to ease the suspension by avoiding the liquid-to-vapor transition, where capillary forces cause the coupons to collapse to the substrate. A first option consists of using a vapor-phase HF tool where the etchant is never in a liquid phase. However, the non-volatile byproduct lithium fluoride (LiF) is redeposited on the top and bottom of the coupon \cite{kaufmann2023redeposition,luke2020wafer}. Alternatively, a liquid HF etchant can be used in combination with critical point drying (CPD), such that any byproducts are rinsed away in the liquid phase. While successful for small area coupons, from a certain size the coupons will collapse due their flexibility, as shown in Fig.\ref{fig:TPchallenges}. While occasionally the edges remain suspended, the coupons themselves collapse. The strong adhesion with the substrate impedes the coupons from being picked or creates significant stress in the thin film during pick-up, leading to crack formation.
Fig.\ref{fig:TPchallenges} depicts such printed coupon on a SiN waveguide, showing cracks through the coupon with non-ideal breaking of the tethers. 

To overcome these issues, we developed a method that relies on pillar-like support structures to keep the LN from collapsing to the substrate, independent of the coupon area. Therefore, the coupons do not adhere to the substrate, avoiding stresses during pick-up and resulting in an excellent transfer yield.

\section{Pillar-based transfer printing method}

\begin{figure}[t]
\centering
\captionsetup{margin=1cm}
\includegraphics[width=\textwidth]{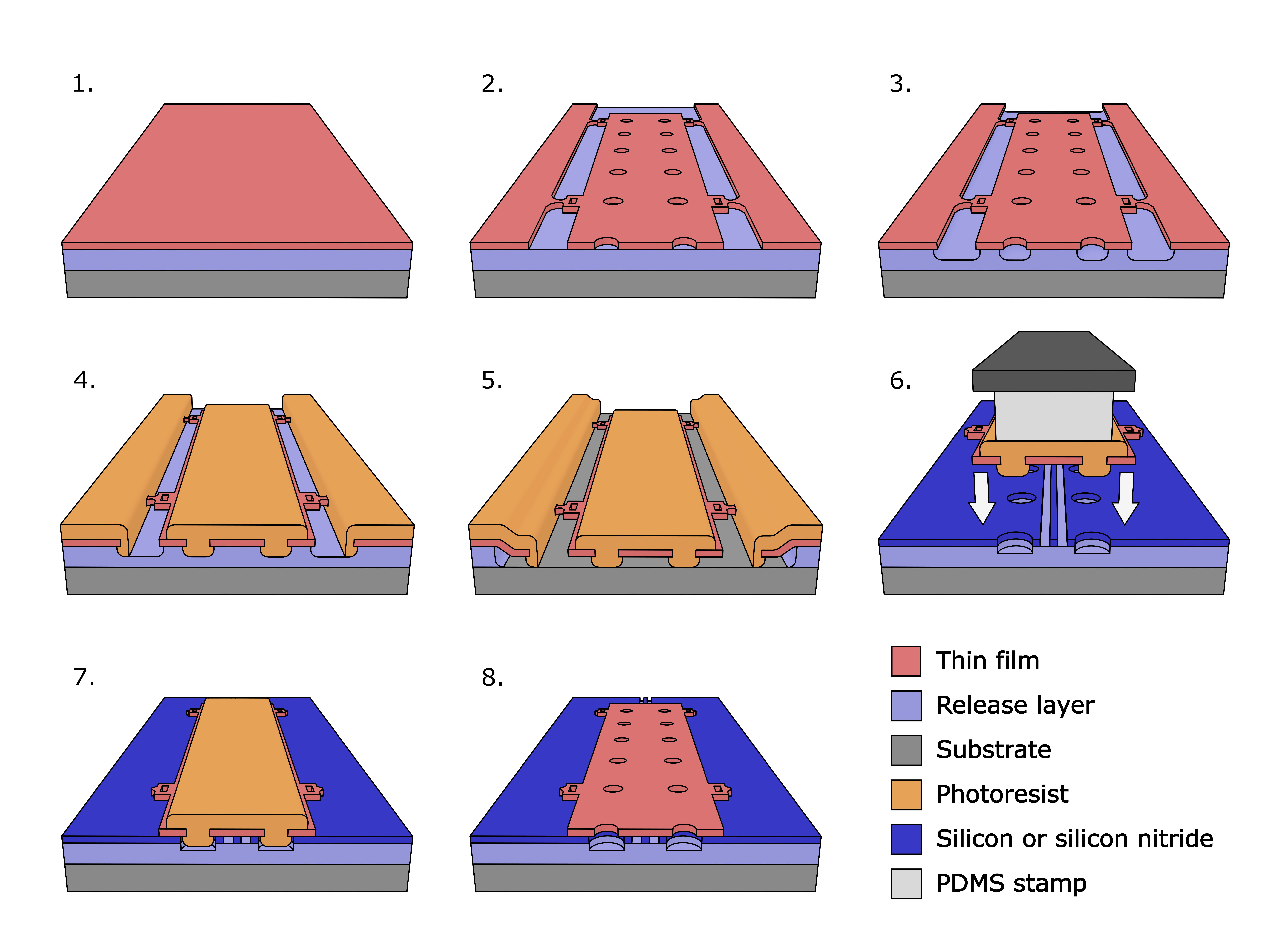}
\caption{Process flow of the pillar-based micro-transfer printing method}
\label{fig:ProcessFlow}
\end{figure}

The process flow is depicted in Fig.\ref{fig:ProcessFlow}. The fabrication starts from a layer stack containing a thin film with a release layer underneath (Fig.\ref{fig:ProcessFlow}.1). For lithium niobate, this consists of 300 nm x-cut LN/2 $\upmu$m SiO$_2$/LN layer stack (NanoLN). A coupon is patterned in the device layer through standard optical lithography and dry etching (Fig.\ref{fig:ProcessFlow}.2). Tethers are defined connecting the coupon to the sides, with holes within the coupon providing access to the release layer underneath. Next, the release layer is partially etched through the holes to create voids that will serve as a mold (Fig.\ref{fig:ProcessFlow}.3). While HF-based wet etching is used, dry etching is also a possibility. These molds are then filled with a support material like photoresist, creating a pillar underneath each patterned hole (Fig.\ref{fig:ProcessFlow}.4). Additionally, covering the coupon with support material increases its mechanical strength. From this point, the coupon is ready to be released from the substrate.

Releasing the coupon is done with an HF-based wet etch, without a CPD step as typical intermediate fluids would dissolve the photoresist. While previously the LN collapsed to the substrate, the thin film is now suspended by the pillars (Fig.\ref{fig:ProcessFlow}.5). Due to the limited contact area between pillars and substrate, the adhesion of the coupon to the substrate is significantly reduced, enabling an easy pick-up with the stamp. In parallel to the source preparation, dedicated recesses are patterned in the target wafer next to the photonic circuit. 
With both source and target wafer prepared, the coupon is picked and printed with the pillars placed within the recesses, such that the thin film bonds directly to the target (Fig.\ref{fig:ProcessFlow}.6 \& \ref{fig:ProcessFlow}.7). Lastly, the photoresist is removed with a standard cleaning procedure, leaving the thin film printed on top of the photonic circuit (Fig.\ref{fig:ProcessFlow}.8).

\section{Optimization of the method}

\begin{figure}[t]
\centering
\captionsetup{margin=1cm}
\includegraphics[width=\textwidth]{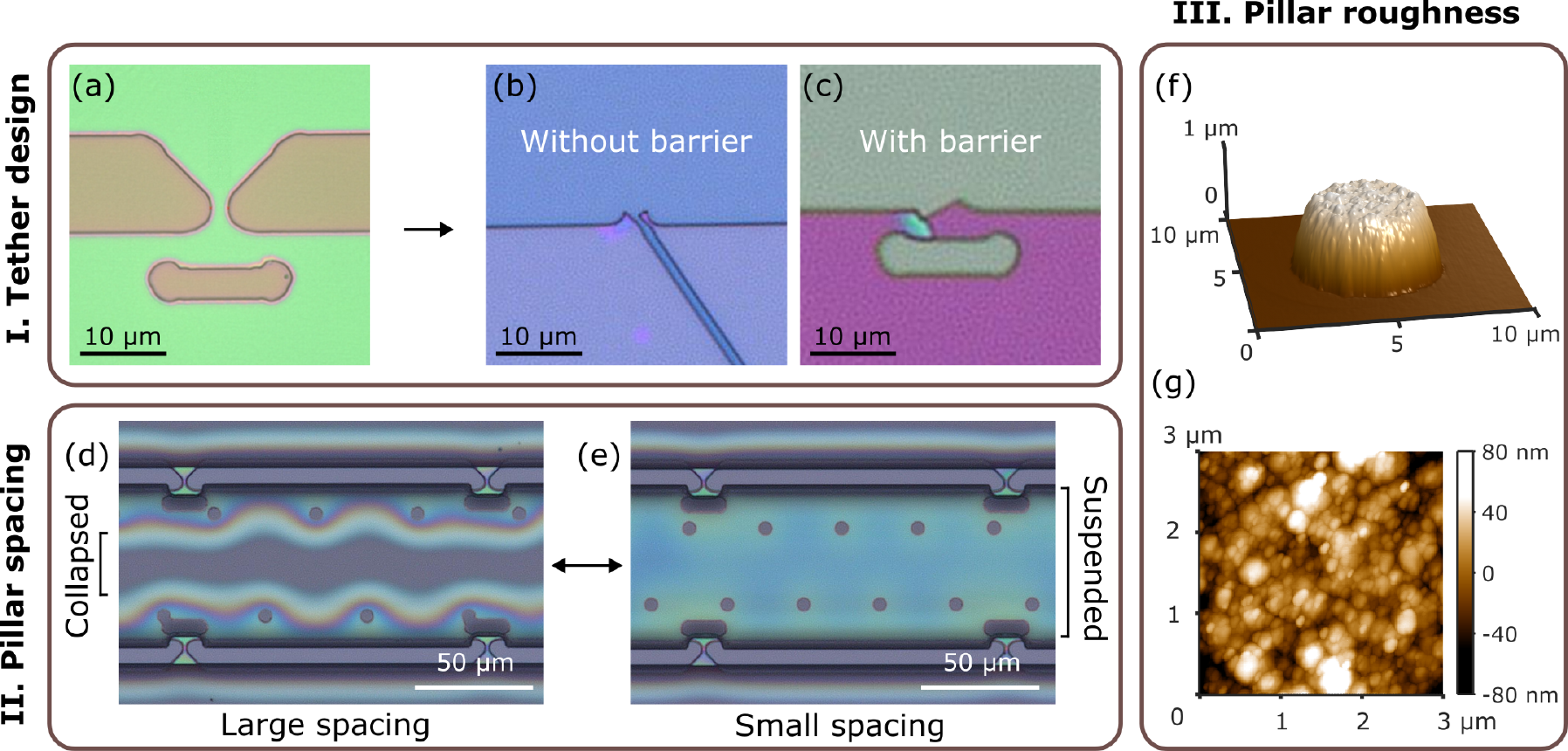}
\caption{Optimization of the pillar-based method: \textbf{I} Optical microscope images of tether design: (a) new tether design with crack barrier, (b) Broken tether without barrier, (c) Broken tether with barrier. \textbf{II} Optical microscope image of pillar spacing effect: (d) Large pillar spacing, (e) Small pillar spacing. \textbf{III} AFM measurements of a pillar: (f) a single pillar underneath the coupon, (g) AFM roughness measurement on top of a pillar.}
\label{fig:Optimizations}
\end{figure}

To ensure a high transfer yield, some further optimizations can be done, as summarized in Fig.\ref{fig:Optimizations}. A first improvement is found in the tether design. A simple hourglass design would be expected to break at its thinnest point. Unfortunately, lithium niobate does not cleave well and sporadically a crack arises that can propagate through the coupon. By adding elongated holes as so-called crack barriers \cite{cuyvers2022high} to the design (Fig.\ref{fig:Optimizations}a), the cracks are shielded and remain localized at the tether. Fig.\ref{fig:Optimizations}b and \ref{fig:Optimizations}c show the effect without and with crack barrier, respectively. Furthermore, in conventional micro-transfer printing methods the tethers must be strong enough to suspend the coupon, while weak enough to break during pick-up. In this method, the tethers should no longer be strong and numerous for suspension, but only serve to keep the coupons attached to the source wafer during the release step. The tether design therefore becomes less critical and a small number suffices.

A further optimization is done by choosing the correct pillar configuration. Due to the sparse layout of typical photonic circuits, pillars and corresponding target recesses can easily be placed around waveguides without putting restrictions on circuit design. Nonetheless, it is important that the pillar spacing is not too large, to avoid collapse between the pillars due to capillary forces, as shown in Fig.\ref{fig:Optimizations}d and \ref{fig:Optimizations}e. Depending on the layer stack and the pillar height, which can be controlled through the partial etch of the release layer, the maximal spacing varies.

A last optimization is found in the pillar definition itself. Although the lithium niobate is no longer in contact with the substrate after release, the pillars themselves are. To guarantee a successful pick-up, the adhesion between pillars and substrate can be minimized by increasing the roughness on top of the pillars. This in itself is determined by the roughness of the mold, i.e. by the partial etch of the release layer in Fig.\ref{fig:ProcessFlow}.3. For the HF wet etch, the roughness can be worsened by increasing the HF concentration and therefore the etch rate. An atomic force microscopy (AFM) measurement on the bottom side of the coupons allows us to image a single pillar, shown in Fig.\ref{fig:Optimizations}f. The resulting rms roughness on top of the pillar is measured to be 24 nm (Fig.\ref{fig:Optimizations}g), which guarantees a bad adhesion with the source substrate.

\section{Micro-transfer printing of lithium niobate}

\begin{figure}[t]
\centering
\captionsetup{margin=1cm}
\includegraphics[width=\textwidth]{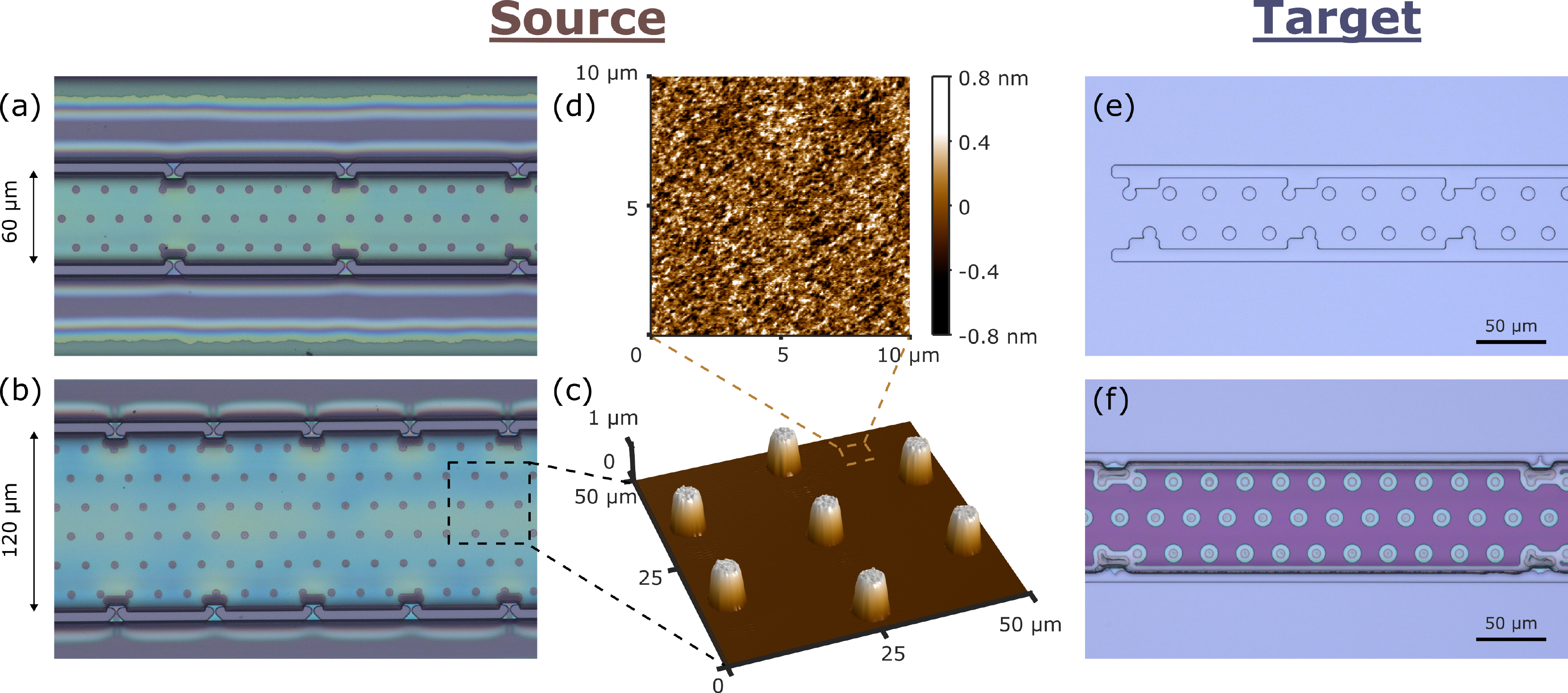}
\caption{Pillar-based $\upmu$TP of LN: (a) Optical microscope image of suspended 60 $\upmu$m wide coupon, (b) Optical microscope image of suspended 120 $\upmu$m wide coupon, (c) AFM measurement of bottom side of a coupon, (d) AFM roughness measurement on bottom side of a coupon, (e) Optical microscope image of a prepared target with pillar recesses, (f) Optical microscope image of a printed LN coupon on the target.}
\label{fig:TPofLN1}
\end{figure}

Having optimized the method, we now demonstrate the micro-transfer printing of 300 nm LN thin films. Fig.\ref{fig:TPofLN1}a shows a 60 $\upmu$m x 1 mm suspended LN coupon on the source chip, ready to be picked. The pillars are configured in a triangular lattice with a spacing of 20 $\upmu$m, where horizontal waveguides can be hosted in between pillars. The green color originates from thin film interference in the air below the coupon, confirming that the lithium niobate is in fact suspended. To demonstrate the area scalability of the method, the same source chip contains 120 $\upmu$m wide coupons, shown in Fig.\ref{fig:TPofLN1}b with the blue color indicating its suspension. This confirms that by adding pillars while maintaining the same pillar spacing, larger coupons are readily suspended and no added or thicker support layers are required. While demonstrated with 120 $\upmu$m wide coupons, the suspended area can be as large as required for high-Q ring or racetrack resonators with sufficiently large bending radii. An AFM measurement reveals the triangular grid of pillars underneath the coupon (Fig.\ref{fig:TPofLN1}c). Moreover, the bottom side of the LN coupon is measured to give an rms roughness of 0.22 nm (Fig.\ref{fig:TPofLN1}d), ideal for direct bonding and low propagation losses. The target chip is prepared by etching pillar recesses deeper than the pillars are high to ensure contact between the coupon and the chip (Fig.\ref{fig:TPofLN1}e). During printing, the pillars are placed within the recesses and LN is directly bonded to the target. Fig.\ref{fig:TPofLN1}f shows a printed coupon, where the uniform purple color indicates a good and uniform direct bonding between coupon and substrate. In a last step, the photoresist is removed, leaving the bonded lithium niobate thin film.\\

\begin{figure}[h]
\centering
\captionsetup{margin=1cm}
\includegraphics[width=\textwidth]{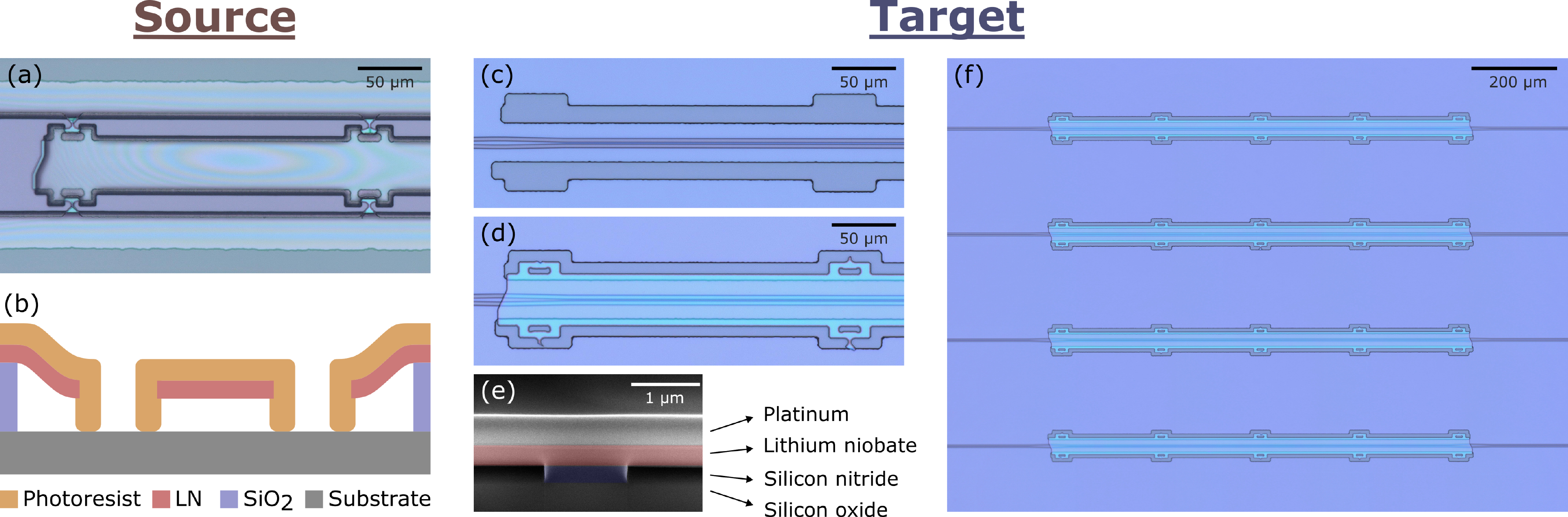}
\caption{$\upmu$TP of LN with a different pillar design: (a) Optical microscope image of suspended LN coupon with side-pillars, (b) Cross section sketch of the LN coupon with side-pillars, (c) Optical microscope image of SiN waveguide surrounded by side-pillar recesses, (d) Optical microscope image of printed LN coupon, (e) False-color SEM cross section of the hybrid LN/SiN waveguide, (f) Optical microscope image of 4 out of 25 subsequently printed LN coupons on SiN waveguides.}
\label{fig:TPofLN2}
\end{figure}

Until now, we relied on a grid of separated pillar-like supports to suspend the coupon. However, the support configuration is defined through optical lithography, which suggests that any kind of support layout can be employed, as long as the maximal spacing is maintained to avoid collapse. For instance, in case the coupon serves to only cover a straight waveguide, long connected pillars along the side of the coupon instead of a grid of separated pillars can be used. Such a suspended coupon is seen in Fig.\ref{fig:TPofLN2}a, with a sketch of its cross section in Fig.\ref{fig:TPofLN2}b. This design freedom allows the pillar configuration to be adapted to the photonic circuit and can be used for stress engineering within the coupon to avoid possible cracks originating from capillary forces during the release etch. With the current pillar design, two long recesses are etched next to a SiN waveguide, as shown in Fig.\ref{fig:TPofLN2}c. Again, the coupon is printed while placing these side-pillars in the corresponding recesses. After removal of the photoresist, the LN coupon is uniformly bonded to the SiN (Fig.\ref{fig:TPofLN2}d), creating a hybrid LN/SiN waveguide.
A SEM image of a corresponding cross section is given in Fig.\ref{fig:TPofLN2}e, with platinum deposited on top for FIB milling. These resulting hybrid waveguides were used to fabricate high-speed modulators with beyond 50 GHz bandwidth \cite{vanackere2023}, and in combination with periodic poling, achieve high-efficiency second harmonic generation with a normalized conversion efficiency of 2500 \%/Wcm$^2$ \cite{vandekerckhove2023}, attesting to their good optical quality. Lastly, to give an estimate of the transfer yield, 25 LN coupons were micro-transfer printed subsequently on top of SiN waveguides without fail, of which four are shown in Fig.\ref{fig:TPofLN2}f. Optical microscope images of all 25 coupons are found in the supplementary material.

\begin{figure}[h]
\centering
\captionsetup{margin=1cm}
\includegraphics[width=0.8\textwidth]{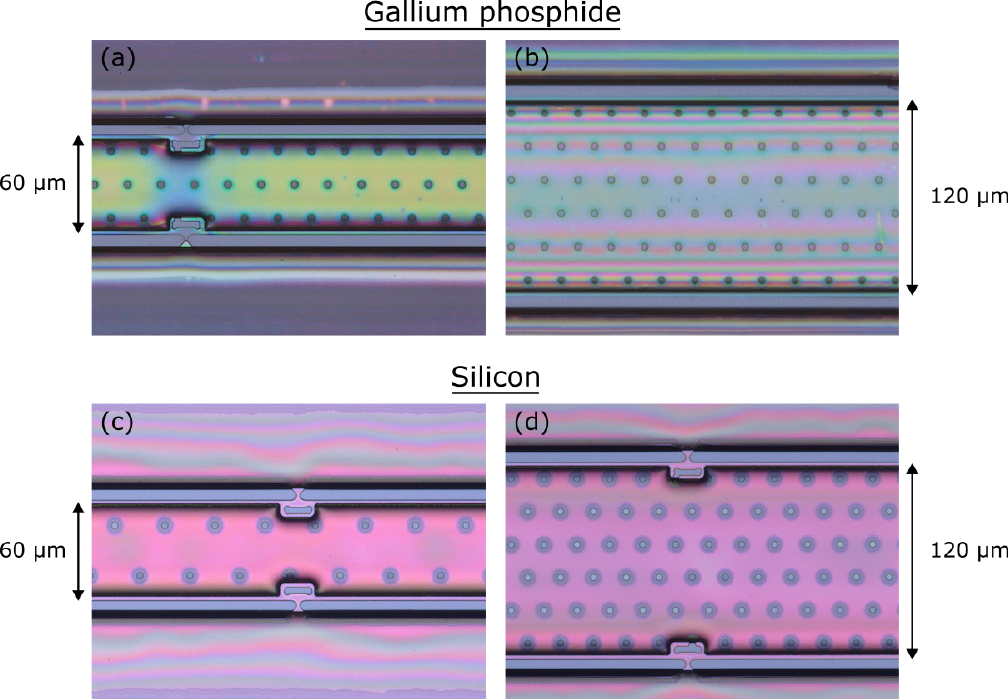}
\caption{$\upmu$TP of gallium phosphide and silicon: optical microscope image of (a) suspended 60 $\upmu$m wide GaP coupon, (b) suspended 120 $\upmu$m wide GaP coupon, (c) suspended 60 $\upmu$m wide Si coupon, and (d) suspended 120 $\upmu$m wide Si coupon.}
\label{fig:Othermaterials}
\end{figure}

\section{Micro-transfer printing of gallium phosphide and silicon}

To demonstrate the versatility of the method, we apply it on other thin-film materials than lithium niobate and fabricate suspended coupons. A first interesting material consists of gallium phosphide, as it possesses a strong $\chi^{(2)}$ and $\chi^{(3)}$ response \cite{wilson2020integrated}. This makes heterogeneous integration onto CMOS-compatible platforms interesting for the same reasons as integrating lithium niobate \cite{billet2022gallium, anthur2021second}. Starting from a layer stack of 300 nm GaP/1 $\upmu$m AlGaP/GaP, the same steps are followed as depicted in Fig.\ref{fig:ProcessFlow}. The intermediate AlGaP layer serves as release layer, as the high aluminium content allows for selective etching compared to the GaP device layer with an HF-based etch. The GaP is patterned into coupons, after which the pillar molds are etched and filled with photoresist. Next, the release layer is etched, resulting in suspended GaP thin films. A 60 $\upmu$m x 1 mm and 120 $\upmu$m x 1 mm sized coupon are shown in Fig.\ref{fig:Othermaterials}a and Fig.\ref{fig:Othermaterials}b, respectively.
While the focus of this work is integration onto CMOS-compatible platforms, integration of single-crystal silicon onto other platforms is also interesting e.g. as intermediate coupling layer on a silicon nitride \cite{poelman2021generic, lufungula2022universally} or lithium-niobate-on-insulator (LNOI) platform \cite{de2021iii, lufungula2022chip}. Starting from a silicon-on-insulator (SOI) chip with a layer stack of 220 nm Si/2 $\upmu$m SiO$_2$/Si, the same procedure is followed. The oxide layer serves as release layer, which can again be selectively etched with an HF-based etch. The resulting suspended coupons are shown in Fig.\ref{fig:Othermaterials}c and Fig.\ref{fig:Othermaterials}d, with a 60 $\upmu$m x 1 mm and 120 $\upmu$m x 1 mm coupon size, respectively. Thus, the same procedure can be followed to readily suspend large areas of both GaP and Si thin films, showing the versatility of the method.

\section{Conclusion}

Micro-transfer printing of thin-film lithium niobate is a promising approach to heterogeneously integrate functionalities that mature CMOS-compatible platforms lack, such as high-speed Pockels modulation and $\chi^{(2)}$ nonlinearities. However, suspending large areas of thin films is challenging, resulting in a low transfer yield. We have developed a new source preparation method to reliably micro-transfer print thin films based on pillar-like support structures. We discussed the optimization of the method by introducing crack barriers in the tether design, controlling the pillar spacing and engineering the pillar roughness. This enables the micro-transfer printing of coupons with varying surface areas, without any additional process development. Furthermore, different support configurations are possible, where pillars along the sides of the coupon were shown for integration onto straight waveguides. An estimate of the transfer yield was given by subsequently printing 25 LN coupons onto SiN waveguides without fail. Moreover, the versatility of the method was demonstrated by fabricating suspended coupons for both gallium phosphide and silicon thin films.
Importantly, this method is complementary to existing LNOI processing. The suspended coupons can contain etched LN waveguides and electrodes, forming optical components that as a whole can be integrated onto a CMOS-compatible platform. Accordingly, this enables the co-integration of LNOI-based high-speed modulators \cite{vanackere2023, yu2022integrated} with integrated lasers \cite{cuyvers2021low} and ultra-fast detectors \cite{maes2023high} on the same chip. By periodically poling the lithium niobate, also $\chi^{(2)}$ processes like second harmonic generation are introduced \cite{vandekerckhove2023, rao2019actively, wang2018ultrahigh}. As such, micro-transfer printing of thin films brings us one step closer to build versatile photonic integrated circuits on CMOS-compatible platforms.\\

\noindent \textbf{Funding.} Research Foundation Flanders (11H6723N, 11F5320N, 11F8122N, 1S69123N), European Research Council (No.759483), Fonds de la Recherche Scientifique (MIS F.4506.20).\\

\noindent \textbf{Acknowledgments.} Tom Vandekerckhove, Tom Vanackere, Stijn Cuyvers and Jasper De Witte are PhD fellows of the Research Foundation Flanders (FWO). Stéphane Clemmen is a research associate of the Fonds de la Recherche Scientifique (FNRS). We also thank the European Research Council (ERC) for the funding in the context of the ELECTRIC project.\\

\noindent \textbf{Supplemental document.} See Supplement 1 for supporting content.\\

\bibliographystyle{ieeetr}
\bibliography{2023OME_pillarmethod}

\begin{thebibliography}{10}

\bibitem{wang2020silicon}
H.~Wang, H.~Chai, Z.~Lv, Z.~Zhang, L.~Meng, X.~Yang, and T.~Yang, ``Silicon
  photonic transceivers for application in data centers,'' {\em Journal of
  Semiconductors}, vol.~41, no.~10, p.~101301, 2020.

\bibitem{rahim2021taking}
A.~Rahim, A.~Hermans, B.~Wohlfeil, D.~Petousi, B.~Kuyken, D.~Van~Thourhout, and
  R.~Baets, ``Taking silicon photonics modulators to a higher performance
  level: state-of-the-art and a review of new technologies,'' {\em Advanced
  Photonics}, vol.~3, no.~2, pp.~024003--024003, 2021.

\bibitem{wang2020integrated}
J.~Wang, F.~Sciarrino, A.~Laing, and M.~G. Thompson, ``Integrated photonic
  quantum technologies,'' {\em Nature Photonics}, vol.~14, no.~5, pp.~273--284,
  2020.

\bibitem{moody20222022}
G.~Moody, V.~J. Sorger, D.~J. Blumenthal, P.~W. Juodawlkis, W.~Loh,
  C.~Sorace-Agaskar, A.~E. Jones, K.~C. Balram, J.~C. Matthews, A.~Laing,
  M.~Davanco, L.~Chang, J.~E. Bowers, N.~Quack, C.~Galland, I.~Aharonovich,
  M.~A. Wolff, C.~Schuck, N.~Sinclair, M.~Loncar, T.~Komljenovic, D.~Weld,
  S.~Mookherjea, S.~Buckley, M.~Radulaski, S.~Reitzenstein, B.~Pingault,
  B.~Machielse, D.~Mukhopadhyay, A.~Akimov, A.~Zheltikov, G.~S~Agarwal,
  K.~Srinivasan, J.~Lu, H.~X. Tang, W.~Jiang, T.~P. McKenna, A.~H.
  Safavi-Naeini, S.~Steinhauer, A.~W. Elshaari, V.~Zwiller, P.~S. Davids,
  N.~Martinez, M.~Gehl, J.~Chiaverini, K.~K. Mehta, J.~Romero, N.~B. Lingaraju,
  A.~M. Weiner, D.~Peace, R.~Cernansky, M.~Lobino, E.~Diamanti,
  L.~Trigo~Vidarte, and R.~M. Camacho, ``2022 roadmap on integrated quantum
  photonics,'' {\em Journal of Physics: Photonics}, vol.~4, no.~1, p.~012501,
  2022.

\bibitem{kaur2021hybrid}
P.~Kaur, A.~Boes, G.~Ren, T.~G. Nguyen, G.~Roelkens, and A.~Mitchell, ``Hybrid
  and heterogeneous photonic integration,'' {\em APL Photonics}, vol.~6, no.~6,
  p.~061102, 2021.

\bibitem{liang2021recent}
D.~Liang and J.~E. Bowers, ``Recent progress in heterogeneous iii-v-on-silicon
  photonic integration,'' {\em Light: Advanced Manufacturing}, vol.~2, no.~1,
  pp.~59--83, 2021.

\bibitem{xiang2021high}
C.~Xiang, J.~Guo, W.~Jin, L.~Wu, J.~Peters, W.~Xie, L.~Chang, B.~Shen, H.~Wang,
  Q.-F. Yang, D.~Kinghorn, M.~Paniccia, K.~J. Vahala, P.~A. Morton, and J.~E.
  Bowers, ``High-performance lasers for fully integrated silicon nitride
  photonics,'' {\em Nature communications}, vol.~12, no.~1, p.~6650, 2021.

\bibitem{de2020heterogeneous}
C.~Op~de Beeck, B.~Haq, L.~Elsinger, A.~Gocalinska, E.~Pelucchi, B.~Corbett,
  G.~Roelkens, and B.~Kuyken, ``Heterogeneous iii-v on silicon nitride
  amplifiers and lasers via microtransfer printing,'' {\em Optica}, vol.~7,
  no.~5, pp.~386--393, 2020.

\bibitem{piels2014low}
M.~Piels, J.~F. Bauters, M.~L. Davenport, M.~J. Heck, and J.~E. Bowers,
  ``Low-loss silicon nitride awg demultiplexer heterogeneously integrated with
  hybrid iii--v/silicon photodetectors,'' {\em Journal of lightwave
  technology}, vol.~32, no.~4, pp.~817--823, 2014.

\bibitem{goyvaerts2020transfer}
J.~Goyvaerts, S.~Kumari, S.~Uvin, J.~Zhang, R.~Baets, A.~Gocalinska,
  E.~Pelucchi, B.~Corbett, and G.~Roelkens, ``Transfer-print integration of
  gaas pin photodiodes onto silicon nitride waveguides for near-infrared
  applications,'' {\em Optics Express}, vol.~28, no.~14, pp.~21275--21285,
  2020.

\bibitem{boynton2020heterogeneously}
N.~Boynton, H.~Cai, M.~Gehl, S.~Arterburn, C.~Dallo, A.~Pomerene, A.~Starbuck,
  D.~Hood, D.~C. Trotter, T.~Friedmann, C.~T. DeRose, and A.~Lentine, ``A
  heterogeneously integrated silicon photonic/lithium niobate travelling wave
  electro-optic modulator,'' {\em Optics express}, vol.~28, no.~2,
  pp.~1868--1884, 2020.

\bibitem{alexander2018nanophotonic}
K.~Alexander, J.~P. George, J.~Verbist, K.~Neyts, B.~Kuyken, D.~Van~Thourhout,
  and J.~Beeckman, ``Nanophotonic pockels modulators on a silicon nitride
  platform,'' {\em Nature communications}, vol.~9, no.~1, pp.~1--6, 2018.

\bibitem{boes2023lithium}
A.~Boes, L.~Chang, C.~Langrock, M.~Yu, M.~Zhang, Q.~Lin, M.~Lon{\v{c}}ar,
  M.~Fejer, J.~Bowers, and A.~Mitchell, ``Lithium niobate photonics: Unlocking
  the electromagnetic spectrum,'' {\em Science}, vol.~379, no.~6627,
  p.~eabj4396, 2023.

\bibitem{ghosh2023wafer}
S.~Ghosh, S.~Yegnanarayanan, D.~Kharas, M.~Ricci, J.~J. Plant, and P.~W.
  Juodawlkis, ``Wafer-scale heterogeneous integration of thin film lithium
  niobate on silicon-nitride photonic integrated circuits with low loss bonding
  interfaces,'' {\em Optics Express}, vol.~31, no.~7, pp.~12005--12015, 2023.

\bibitem{snigirev2023ultrafast}
V.~Snigirev, A.~Riedhauser, G.~Lihachev, M.~Churaev, J.~Riemensberger, R.~N.
  Wang, A.~Siddharth, G.~Huang, C.~M{\"o}hl, Y.~Popoff, U.~Drechsler, D.~Caimi,
  S.~Honl, J.~Liu, P.~Seidler, and T.~J. Kippenberg, ``Ultrafast tunable lasers
  using lithium niobate integrated photonics,'' {\em Nature}, vol.~615,
  no.~7952, pp.~411--417, 2023.

\bibitem{vanackere2020micro}
T.~Vanackere, M.~Billet, C.~Op~de Beeck, S.~Poelman, G.~Roelkens, S.~Clemmen,
  and B.~Kuyken, ``Micro-transfer printing of lithium niobate on silicon
  nitride,'' in {\em 2020 European Conference on Optical Communications
  (ECOC)}, pp.~1--4, IEEE, 2020.

\bibitem{li2022photonic}
Z.~Li, J.~A. Smith, M.~Scullion, N.~K. Wessling, L.~J. McKnight, M.~D. Dawson,
  and M.~J. Strain, ``Photonic integration of lithium niobate micro-ring
  resonators onto silicon nitride waveguide chips by transfer-printing,'' {\em
  Optical Materials Express}, vol.~12, no.~11, pp.~4375--4383, 2022.

\bibitem{roelkens2022micro}
G.~Roelkens, J.~Zhang, L.~Bogaert, M.~Billet, D.~Wang, B.~Pan, C.~J. Kruckel,
  E.~Soltanian, D.~Maes, T.~Vanackere, T.~Vandekerckhove, S.~Cuyvers,
  J.~De~Witte, I.~Luntadila~Lufungula, X.~Guo, H.~Li, S.~Qin, G.~Muliuk,
  S.~Uvin, B.~Haq, C.~Op~de Beeck, J.~Goyvaerts, G.~Lepage, P.~Verheyen,
  J.~Van~Campenhout, G.~Morthier, B.~Kuyken, D.~Van~Thourhout, and R.~Baets,
  ``Micro-transfer printing for heterogeneous si photonic integrated
  circuits,'' {\em IEEE Journal of Selected Topics in Quantum Electronics},
  vol.~29, no.~3: Photon. Elec. Co-Inte. and Adv. Trans. Print., pp.~1--14,
  2022.

\bibitem{kaufmann2023redeposition}
F.~Kaufmann, G.~Finco, A.~Maeder, and R.~Grange, ``Redeposition-free
  inductively-coupled plasma etching of lithium niobate for integrated
  photonics,'' {\em Nanophotonics}, 2023.

\bibitem{luke2020wafer}
K.~Luke, P.~Kharel, C.~Reimer, L.~He, M.~Loncar, and M.~Zhang, ``Wafer-scale
  low-loss lithium niobate photonic integrated circuits,'' {\em Optics
  Express}, vol.~28, no.~17, pp.~24452--24458, 2020.

\bibitem{cuyvers2022high}
S.~Cuyvers, T.~Vanackere, T.~Vandekerckhove, S.~Poelman, C.~Op~de Beeck,
  J.~De~Witte, A.~Hermans, K.~Van~Gasse, N.~Picqu{\'e}, D.~Van~Thourhout,
  G.~Roelkens, S.~Clemmen, and B.~Kuyken, ``High-yield heterogeneous
  integration of silicon and lithium niobate thin films,'' in {\em 2022
  Conference on Lasers and Electro-Optics (CLEO)}, pp.~1--2, IEEE, 2022.

\bibitem{vanackere2023}
T.~Vanackere, T.~Vandekerckhove, L.~Bogaert, M.~Billet, S.~Poelman,
  J.~Van~Kerrebrouck, A.~Moerman, O.~Caytan, S.~Lemey, G.~Torfs, G.~Roelkens,
  S.~Clemmen, and B.~Kuyken, ``High-speed lithium niobate modulator on silicon
  nitride using micro-transfer printing,'' in {\em 2023 Conference on Lasers
  and Electro-Optics (CLEO)}, pp.~1--2, IEEE, 2023.
\newblock (to be published).

\bibitem{vandekerckhove2023}
T.~Vandekerckhove, T.~Vanackere, J.~De~Witte, I.~Luntadila~Lufungula,
  E.~Vissers, G.~Roelkens, S.~Clemmen, and B.~Kuyken, ``High-efficiency second
  harmonic generation in heterogeneously-integrated periodically poled lithium
  niobate on silicon nitride,'' in {\em 2023 Conference on Lasers and
  Electro-Optics Europe \& European Quantum Electronics Conference
  (CLEO/Europe-EQEC)}, pp.~1--1, IEEE, 2023.
\newblock (to be published).

\bibitem{wilson2020integrated}
D.~J. Wilson, K.~Schneider, S.~H{\"o}nl, M.~Anderson, Y.~Baumgartner,
  L.~Czornomaz, T.~J. Kippenberg, and P.~Seidler, ``Integrated gallium
  phosphide nonlinear photonics,'' {\em Nature Photonics}, vol.~14, no.~1,
  pp.~57--62, 2020.

\bibitem{billet2022gallium}
M.~Billet, L.~Reis, Y.~L{\'e}ger, C.~Cornet, F.~Raineri, I.~Sagnes, K.~Pantzas,
  G.~Beaudoin, G.~Roelkens, F.~Leo, and B.~Kuyken, ``Gallium
  phosphide-on-insulator integrated photonic structures fabricated using
  micro-transfer printing,'' {\em Optical Materials Express}, vol.~12, no.~9,
  pp.~3731--3737, 2022.

\bibitem{anthur2021second}
A.~P. Anthur, H.~Zhang, Y.~Akimov, J.~R. Ong, D.~Kalashnikov, A.~I. Kuznetsov,
  and L.~Krivitsky, ``Second harmonic generation in gallium phosphide
  nano-waveguides,'' {\em Optics Express}, vol.~29, no.~7, pp.~10307--10320,
  2021.

\bibitem{poelman2021generic}
S.~Poelman, S.~Cuyvers, J.~De~Witte, A.~Hermans, K.~Van~Gasse, N.~Picqu{\'e},
  G.~Roelkens, D.~Van~Thourhout, and B.~Kuyken, ``Generic heterogeneous
  integration process flow for commercial foundry low-index photonic
  platforms,'' in {\em Frontiers in Optics}, pp.~FM1B--6, Optica Publishing
  Group, 2021.

\bibitem{lufungula2022universally}
I.~L. Lufungula, A.~Shams-Ansari, D.~Renaud, C.~Op~de Beeck, S.~Cuyvers,
  S.~Poelman, G.~Roelkens, M.~Loncar, and B.~Kuyken, ``Universally printable
  single-mode laser on low-index platforms,'' in {\em 2022 Conference on Lasers
  and Electro-Optics (CLEO)}, pp.~1--2, IEEE, 2022.

\bibitem{de2021iii}
C.~Op~de Beeck, F.~M. Mayor, S.~Cuyvers, S.~Poelman, J.~F. Herrmann, O.~Atalar,
  T.~P. McKenna, B.~Haq, W.~Jiang, J.~D. Witmer, G.~Roelkens,
  A.~H.~Safavi-Naeini, R.~Van~Laer, and B.~Kuyken, ``Iii/v-on-lithium niobate
  amplifiers and lasers,'' {\em Optica}, vol.~8, no.~10, pp.~1288--1289, 2021.

\bibitem{lufungula2022chip}
I.~L. Lufungula, A.~Shams-Ansari, D.~Renaud, C.~Op~de Beeck, S.~Cuyvers,
  S.~Poelman, G.~Roelkens, M.~Loncar, and B.~Kuyken, ``On-chip electro-optic
  frequency comb generation using a heterogeneously integrated laser,'' in {\em
  CLEO: QELS\_Fundamental Science}, pp.~JTh6B--7, Optica Publishing Group,
  2022.

\bibitem{yu2022integrated}
M.~Yu, D.~Barton~III, R.~Cheng, C.~Reimer, P.~Kharel, L.~He, L.~Shao, D.~Zhu,
  Y.~Hu, H.~R. Grant, L.~Johansson, Y.~Okawachi, A.~L. Gaeta, M.~Zhang, and
  M.~Loncar, ``Integrated femtosecond pulse generator on thin-film lithium
  niobate,'' {\em Nature}, pp.~1--7, 2022.

\bibitem{cuyvers2021low}
S.~Cuyvers, B.~Haq, C.~Op~de Beeck, S.~Poelman, A.~Hermans, Z.~Wang,
  A.~Gocalinska, E.~Pelucchi, B.~Corbett, G.~Roelkens, K.~Van~Gasse, and
  B.~Kuyken, ``Low noise heterogeneous iii-v-on-silicon-nitride mode-locked
  comb laser,'' {\em Laser \& Photonics Reviews}, vol.~15, no.~8, p.~2000485,
  2021.

\bibitem{maes2023high}
D.~Maes, S.~Lemey, G.~Roelkens, M.~Zaknoune, V.~Avramovic, E.~Okada,
  P.~Szriftgiser, E.~Peytavit, G.~Ducournau, and B.~Kuyken, ``High-speed
  uni-traveling-carrier photodiodes on silicon nitride,'' {\em APL Photonics},
  vol.~8, no.~1, p.~016104, 2023.

\bibitem{rao2019actively}
A.~Rao, K.~Abdelsalam, T.~Sjaardema, A.~Honardoost, G.~F. Camacho-Gonzalez, and
  S.~Fathpour, ``Actively-monitored periodic-poling in thin-film lithium
  niobate photonic waveguides with ultrahigh nonlinear conversion efficiency of
  4600\% w- 1 cm- 2,'' {\em Optics express}, vol.~27, no.~18, pp.~25920--25930,
  2019.

\bibitem{wang2018ultrahigh}
C.~Wang, C.~Langrock, A.~Marandi, M.~Jankowski, M.~Zhang, B.~Desiatov, M.~M.
  Fejer, and M.~Lon{\v{c}}ar, ``Ultrahigh-efficiency wavelength conversion in
  nanophotonic periodically poled lithium niobate waveguides,'' {\em Optica},
  vol.~5, no.~11, pp.~1438--1441, 2018.

\end{thebibliography}

\end{document}


\maketitle

\begin{abstract}
In the main manuscript, the claim is made that 25 lithium niobate coupons are micro-transfer printed without fail to demonstrate the reliability of the method. This supplementary material serves to provide evidence for this claim by showing optical microscope images of all 25 printed coupons.
\end{abstract}

\section{Optical microscope images of 25 printed lithium niobate coupons}

\begin{figure}[h!]
\centering
\includegraphics[width=\textwidth]{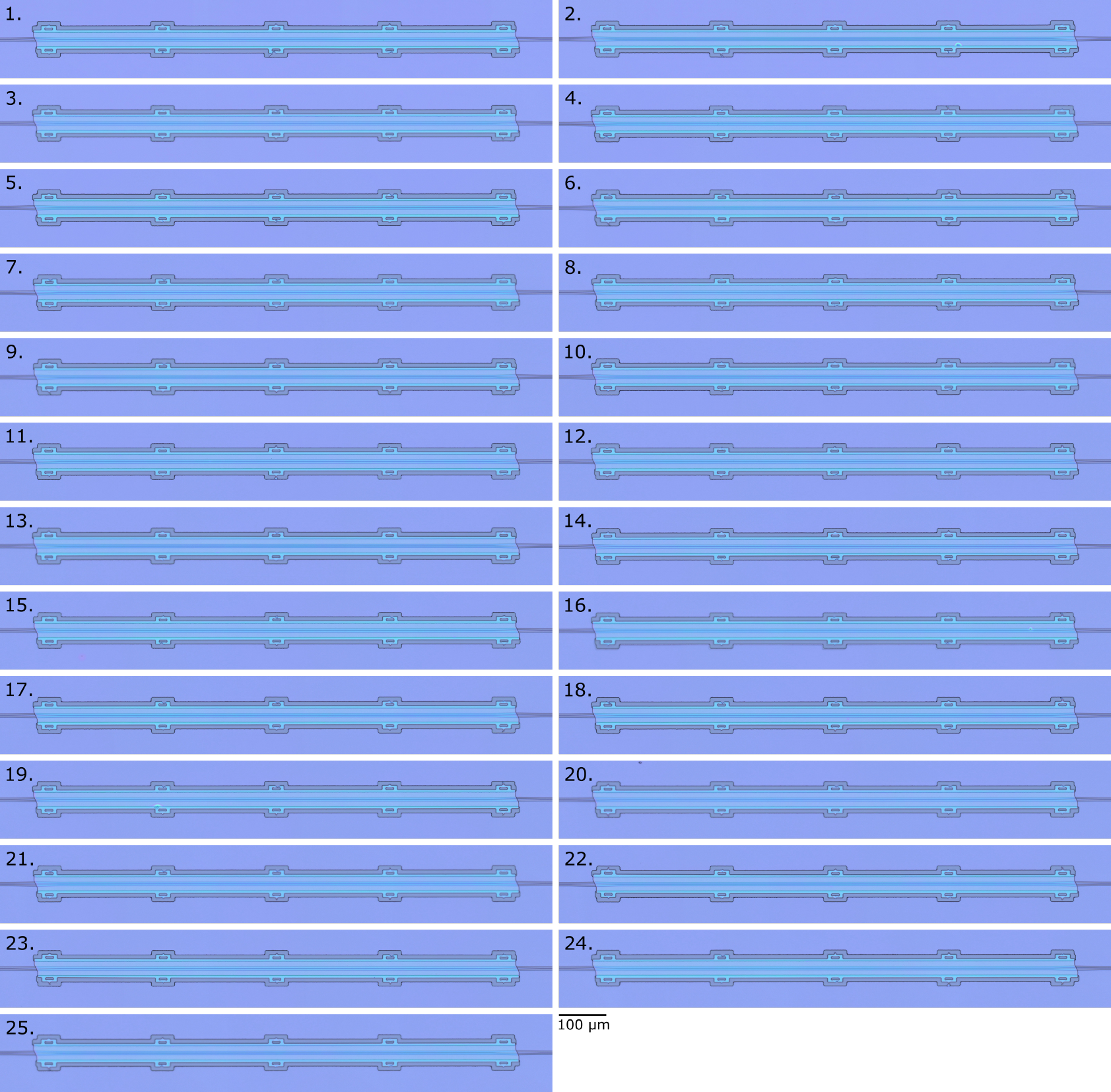}
\caption{25 micro-transfer printed lithium niobate coupons onto silicon nitride waveguides.}
\label{fig:25coupons}
\end{figure}

The 25 lithium niobate coupons on silicon nitride waveguides are shown in Fig.\ref{fig:25coupons}. 

